\documentclass[lettersize,journal]{IEEEtran}
\usepackage{algorithmic}
\usepackage{algorithm}
\usepackage{array}
\usepackage{soul}
\usepackage{amsmath,amsfonts,amssymb}
\usepackage[caption=false,font=normalsize,labelfont=sf,textfont=sf]{subfig}
\usepackage{textcomp}
\usepackage{stfloats}
\usepackage{url}
\usepackage{verbatim}
\usepackage{graphicx}
\usepackage[style=ieee,backend=bibtex]{biblatex}
\DeclareFieldFormat{doi}{} %de-display doi
\usepackage{cleveref}
\crefname{figure}{Fig.}{Figs.}
\Crefname{figure}{Fig.}{Figs.}
\crefname{equation}{Eq.}{Eqs.}
\Crefname{equation}{Eq.}{Eqs.}
\usepackage{xcolor}

\DeclareFieldFormat{doi}{} %de-display doi
\usepackage{titlesec}

\titlespacing*{\section}{0pt}{0.8ex plus 0.2ex minus 0.2ex}{0.4ex plus 0.2ex}
\titlespacing*{\subsection}{0pt}{0.6ex plus 0.2ex minus 0.2ex}{0.3ex plus 0.1ex}
\titlespacing*{\subsubsection}{0pt}{0.5ex plus 0.2ex minus 0.2ex}{0.25ex plus 0.1ex}

\begin{document}

\title{Compressive Beam-Pattern–Aware Near-field Beam Training via Total Variation Denoising}

\author{Zijun Wang, Maria Nivetha A, Ye Hu,~\IEEEmembership{Member,~IEEE}, Rui Zhang,~\IEEEmembership{Member,~IEEE}
        % <-this % stops a space
\thanks{Zijun Wang, Maria Nivetha A and Rui Zhang are with the Department of Electrical Engineering, The State University of New York at Buffalo, New York, USA  (email: {zwang267@buffalo.edu},{marianiv@buffalo.edu},{rzhang45@buffalo.edu}). Ye Hu is with the Industrial and Systems Engineering, and Department of Electrical and Computer Engineering, University of Miami, Florida, USA (email: {yehu@miami.edu}). }% <-this % stops a space
\thanks{This work was supported in part by the CSD, MediaTek Inc. USA, under Grant 103764, and in part by the National Science Foundation, under Grant ECCS 2512911.}}

% The paper headers
\markboth{Journal of \LaTeX\ Class Files,~Vol.~14, No.~8, August~2021}%
{Shell \MakeLowercase{\textit{et al.}}: A Sample Article Using IEEEtran.cls for IEEE Journals}

\IEEEpubid{0000--0000/00\$00.00~\copyright~2021 IEEE}
% Remember, if you use this you must call \IEEEpubidadjcol in the second
% column for its text to clear the IEEEpubid mark.

\maketitle

\begin{abstract}
Extremely large antenna arrays envisioned for 6G incurs near-field effect, where steering vector depends on angles and range simultaneously. Polar-domain near-field codebooks can focus energy accurately but incur extra two-dimensional sweeping overhead; compressed-sensing (CS) approaches with Gaussian-masked DFT sensing offer a lower-overhead alternative. This letter revisits near-field beam training using conventional DFT codebooks. Unlike far-field responses that concentrate energy on a few isolated DFT beams, near-field responses produce contiguous, plateau-like energy segments with sharp transitions in the DFT beamspace. Pure LASSO denoising, therefore, tends to over-shrink magnitudes and fragment plateaus. We propose a beam-pattern–preserving beam training scheme for {multiple-path scenarios} that combines LASSO with a lightweight denoising pipeline: LASSO to suppress small-amplitude noise, followed by total variation (TV) to maintain plateau levels and edge sharpness. The two proximal steps require no near-field codebook design. Simulations with Gaussian pilots show consistent NMSE and cosine-similarity gains over least squares and LASSO at the same pilot budget.
\end{abstract}


\section{Introduction}

\IEEEPARstart{T}{he} imminent 6G era will exploit a broadened spectrum portfolio
that extends beyond traditional sub-6~GHz bands to the emerging upper mid-band/FR3
(7--24~GHz), while also leveraging mmWave and sub-THz bands to support multi-Gbps
connectivity and new services \cite{miao_fr3_survey_arxiv2501,kang_uppermidband_ojcoms24,zhang_newmidband_mcom25}.
To compensate for increased propagation loss at higher frequencies, extremely
large antenna arrays (ELAA) with hundreds to thousands of elements are expected,
which incur the radiative near-field effect in wireless networks
\cite{lu_nearfield_xlmimo_comst24}.
Unlike the far-field (i.e., plane-wave) model, where steering vectors depends only on angle, the near-field channel model exhibits spherical wavefronts and steering vectors that depend
on both angle and range \cite{channel_model2,General_tutorial}.

This shift motivated extensive research on channel modeling, beam training, and channel estimation in the near-field region \cite{Cui_and_Dai_channel_model,wuMultipleAccessNearField2023}. Some works directly extend conventional far-field beam-training schemes to near-field line-of-sight (LoS) links \cite{Fast_Near-Field_Beam_Training,Joint_Angle_and_Range_Estimation}. 
\cite{Fast_Near-Field_Beam_Training} estimates the angle via a DFT codebook and then performs a range search conditioned on the estimated direction, whereas \cite{Joint_Angle_and_Range_Estimation} uses DFT sweeping and jointly estimates angle and range from received power measurements.
Moreover, \cite{Cui_and_Dai_channel_model} develops a polar–domain codebook that embeds both angular and range information for uplink channel estimation. While such polar-domain codebooks enable accurate near-field beam focusing, their two-dimensional sampling over the distance and angle domain inevitably enlarges the beam-sweeping set and thus inflates the training and the storage overhead. Also, such codebook is not supported by the angle-only CSI codebook framework in current standard \cite{qin2023reviewcodebookscsifeedback}. Motivated by conventional far-field channel estimation, it is natural instead to regard the near-field channel as sparse in a suitable transform domain and adopt compressive sensing (CS) to estimate the corresponding coefficients. Classical CS theory shows that random-masked Fourier operators provide stable recovery via LASSO under Restricted Isometry Property (RIP) and related incoherence conditions \cite{donoho2006compressed,candes2005decoding}. 
For example, \cite{lee_tcomm16_omp_hybrid_ce} formulates channel estimation as a sparse recovery problem and applies orthogonal matching pursuit (OMP) to identify the dominant AoA/AoD components under hybrid beamforming constraints.
More generally, \cite{mendezrial_access16_hybrid_arch_cs_ce} proposes a compressive channel estimation framework applicable across different hybrid hardware realizations.
In contrast, sensing built from highly focused near-field codewords is typically much more coherent and thus lacks RIP-type guarantees \cite{rudelson2008fourier}.

% Classical CS theory shows that some random-masked Fourier operators provide stable recovery via LASSO under the Restricted Isometry Property (RIP) and related incoherence conditions\cite{donoho2006compressed,candes2005decoding}. This has been widely adopted in far-field beam training.
% For example, \cite{lee_tcomm16_omp_hybrid_ce} formulates channel estimation as a sparse recovery problem and applies orthogonal matching pursuit (OMP) to identify the dominant AoA/AoD components under hybrid beamforming constraints.
% More generally, \cite{mendezrial_access16_hybrid_arch_cs_ce} proposes a compressive channel estimation framework applicable across different hybrid hardware realizations.
% In contrast, highly focused near-field codebooks tend to exhibit large mutual coherence and typically fail to satisfy RIP-type guarantees\cite{rudelson2008fourier}. 

\IEEEpubidadjcol
\vspace{-1mm}
{Because of the mismatch between polar-domain near-field codebooks and classical compressed sensing}, we explore using conventional DFT codebooks that are typically used in far-field users as a low-overhead CS approach to near-field beam training.
%The incompatibility between polar-domain near-field codebooks and classical compressed sensing motivates us to revisit DFT codebooks used in far-field conditions for near-field beam training within a CS framework.
{Near-field effects cause energy to split across neighboring DFT beams} \cite{wang2025lowcomplexitynearfieldbeamtraining,Cui_and_Dai_channel_model}, but when scattering is limited (e.g., higher frequency band), the channel can still be accurately captured by a small number of DFT codewords, preserving sparsity that CS methods exploit \cite{wang2025sparsityawarenearfieldbeamtraining}. Consequently, near-field DFT-beamspace responses form contiguous, plateau-like regions rather than isolated peaks\cite{wang2025precisenearfieldbeamtraining}. Standard LASSO regularization commonly over-shrinks coefficient magnitudes and encourages isolated nonzeros, which can break up or attenuate such plateaus. Total variation (TV) denoising, however, encourages piecewise-constant solutions and preserves sharp transitions, so combining LASSO with TV naturally suits beamspace patterns that are both sparse and plateau-structured \cite{chambolle2004tv}.

Motivated by these insights, this letter proposes a low-overhead compressive training framework that pairs Gaussian-masked DFT sensing with a two-stage denoising pipeline: LASSO to suppress small-amplitude noise, followed by TV denoising to preserve plateau heights and sharp transitions. The proposed approach requires no specialized near-field codebook and is applicable to both ULA (uniform linear array) and UPA (uniform planar array) configurations. Importantly, our method can be implemented directly on top of existing DFT-based codebook architectures, enabling compatibility with current standard-compliant transceivers \cite{qin2023reviewcodebookscsifeedback}.

The rest of the paper is organized as follows. \Cref{sec:system} introduces the near-field system model for a UPA. \Cref{sec:beampattern} analyzes the DFT beam patterns and the resulting sparsity structure. \Cref{sec:ulamethod} presents the proposed compressive beam training algorithm. \Cref{sec:upa_results} reports simulation results for the UPA scenario. Finally, \Cref{sec:clu} concludes the paper.

\section{System Model}\label{sec:system}
\begin{figure}[h]
\captionsetup{justification=justified,singlelinecheck=false} 
\centering
\includegraphics[width=0.8\linewidth]{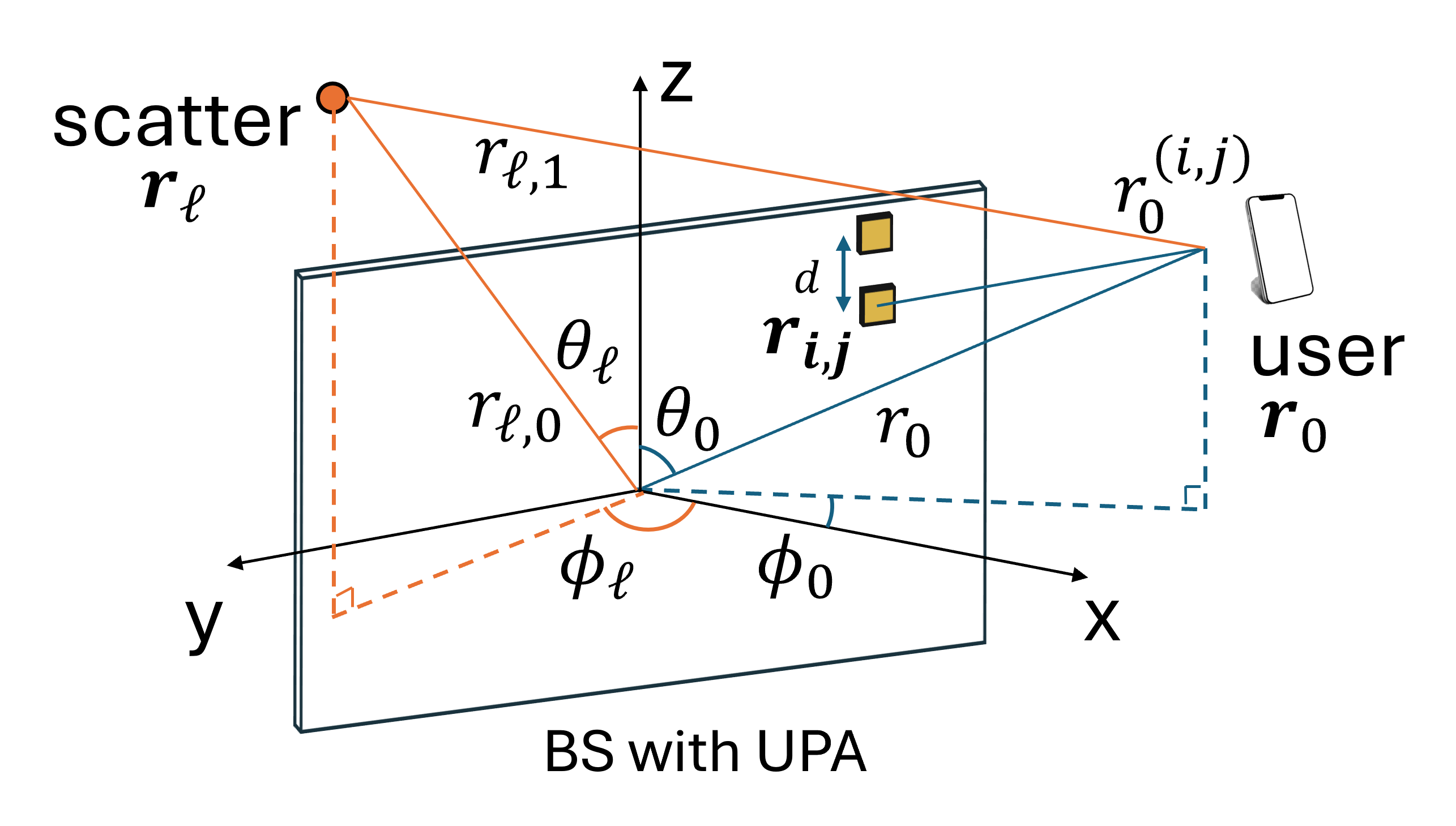}
\caption{Diagram of the near-field communication system featuring a UPA.}
\label{fig:antenna}
\end{figure}

\noindent\hspace*{1em} We consider a downlink narrowband system where a base station (BS) is
equipped with an $N_y\times N_z$ uniform planar array (UPA) serving a
single-antenna user. The UPA lies on the $y$--$z$ plane and is centered
at the origin. Each user terminal
is equipped with a single antenna. The carrier frequency is denoted by
$f_c$ and the corresponding wavelength is $\lambda = c/f_c$, with $c$
being the speed of light. We focus on the radiative near-field regime where the user range lies between the Fresnel distance $r_{\mathrm{F}}$ and the Rayleigh distance $r_{\mathrm{R}}$ \cite{wuMultipleAccessNearField2023}. The whole system is shown in \Cref{fig:antenna}.

Let
$
\mathcal{N}_y \triangleq \{0,1,\dots,N_y-1\},\quad
\mathcal{N}_z \triangleq \{0,1,\dots,N_z-1\},
$
and define centered index offsets
$
\delta_i \triangleq \frac{2i-N_y+1}{2},\;
\delta_j \triangleq \frac{2j-N_z+1}{2},
\; i\in\mathcal{N}_y,\; j\in\mathcal{N}_z.
$
The $(i,j)$-th antenna element is located at $
\mathbf{r}_{i,j} = (0,\,\delta_i d,\,\delta_j d)^{\mathsf{T}},
$
where the inter-element spacing is $d=\lambda/2$ and the total number of
antennas is $N \triangleq N_yN_z$.

A point in space (i.e., user or scatterer) is described in spherical
coordinates by range $r>0$, elevation angle $\theta\in[0,\pi]$ and azimuth angle
$\phi\in[-\pi,\pi]$. We adopt the unit direction vector
\begin{equation}
\mathbf{q}(\theta,\phi) \triangleq
\begin{bmatrix}
\sin\theta\cos\phi\\[0.5mm]
\sin\theta\sin\phi\\[0.5mm]
\cos\theta
\end{bmatrix},
\end{equation}
so that the Cartesian position of a point at $(\theta,\phi,r)$ is
$r\,\mathbf{q}(\theta,\phi)$. The distance between this point and the
$(i,j)$-th array element is
$
r^{(i,j)}(\theta,\phi,r)
= \big\|r\,\mathbf{q}(\theta,\phi)-\mathbf{r}_{i,j}\big\|_2.
$
The corresponding near-field steering response of this path on the UPA
is
\begin{equation}
\big[b(\theta,\phi,r)\big]_{i,j}
= \frac{1}{\sqrt{N}}
\exp\!\left(
 -j\frac{2\pi}{\lambda}
 \big(r^{(i,j)}(\theta,\phi,r)-r\big)
\right),
\label{eq:upa_steering_entry_compact}
\end{equation}
and stacking all $(i,j)$ into a fixed order yields the steering vector
$\mathbf{b}(\theta,\phi,r)\in\mathbb{C}^{N}$.

We assume a single-antenna user is located at
$(\theta_0,\phi_0,r_0)$, where $r_0$ is the distance between user and BS. The
LoS component is modeled as
\begin{equation}
\mathbf{h}_0
= \sqrt{N}\,g_0
  e^{-j\frac{2\pi r_0}{\lambda}}
  \mathbf{b}(\theta_0,\phi_0,r_0),
\label{eq:upa_channel_los_r0}
\end{equation}
with large-scale gain $g_0 = \frac{\lambda}{4\pi r_0}.$

In addition, we consider $(L-1)$ single-bounce scatterers indexed by
$\ell=1,\dots,L-1$. The $\ell$-th scatterer is located at
$(\theta_\ell,\phi_\ell,r_{\ell,0})$ with respect to the BS, so its
position vector is
$
\mathbf{r}_\ell \triangleq
r_{\ell,0}\,\mathbf{q}(\theta_\ell,\phi_\ell).
$
The BS--scatter distance is therefore $r_{\ell,0}=\|\mathbf{r}_\ell\|_2$,
and the scatter--user distance is
$
r_{\ell,1}
= \big\|r_0\,\mathbf{q}(\theta_0,\phi_0)
      - r_{\ell,0}\,\mathbf{q}(\theta_\ell,\phi_\ell)\big\|_2.
$
The NLoS contribution of the $\ell$-th path is modeled as
\begin{equation}
\mathbf{h}_\ell
= \sqrt{N}\,g_\ell
  e^{-j\frac{2\pi (r_{\ell,0}+r_{\ell,1})}{\lambda}}
  \mathbf{b}(\theta_\ell,\phi_\ell,r_{\ell,0}),
\label{eq:upa_channel_each_nlos_r_l0_l1}
\end{equation}
where $g_\ell
= \frac{\lambda}{4\pi r_{\ell,0} r_{\ell,1}}\,p_\ell,
\;
p_\ell \sim \mathcal{CN}(0,1),$
captures both the two-hop path loss and the random complex reflection
coefficient of the $\ell$-th scatterer.
Collecting the LoS and NLoS contributions, the overall downlink channel is
\begin{equation}
\mathbf{h}
= \mathbf{h}_0 + \sum_{\ell=1}^{L-1}\mathbf{h}_\ell
\in\mathbb{C}^{N}.
\label{eq:upa_channel_total_compact}
\end{equation}

% Based on the channel model, the received signal $y(\mathbf{v})$ at a user located at $(\theta, r)$ is expressed as:
% \begin{equation}
%   y(\mathbf{v})=\mathbf{h}^{H}\mathbf{v} x+w,  
%   \label{eq:received signal model}
% \end{equation}
% where $\mathbf{v}$ represents the beamforming vector, and $w\sim \mathcal{CN}(0,\sigma^2)$ is the complex additive white Gaussian noise (AWGN). The reference pilot signal is denoted as $x=1$ without loss of generality.

\section{Analysis of beam pattern}\label{sec:beampattern}
\noindent\hspace*{1em} This section characterizes how a near-field path appears in the 2D DFT
beamspace of a UPA. In contrast to the far-field case where energy concentrated
on a single DFT index, near-field Fresnel phases spread the energy over a
connected neighborhood, forming plateau-like regions with sharp
edges. We first show that, under a separable Fresnel approximation, the 2D
beam pattern of UPA factorizes into two 1D uniform linear array (ULA) patterns, and then use the 6-dB
lobe width law of ULA in \cite{wang2025lowcomplexitynearfieldbeamtraining} to
quantify the per-axis spreading widths and the resulting effective sparsity.

We define the two DFT spatial variables
$
u \triangleq \sin\theta\,\sin\phi,
v \triangleq \cos\theta.
$
With this parameterization, the 2D DFT sampling that ensures
orthogonality corresponds to uniform grids on $(u,v)$, i.e.,
$
u_{n}=\frac{2n-N_y+1}{N_y},
v_{m}=\frac{2m-N_z+1}{N_z},
$
for DFT indices $n\in\{0,\dots,N_y-1\}$ and $m\in\{0,\dots,N_z-1\}$.

We adopt the standard DFT codebook as the beamspace basis on the
UPA. Along the $y$-axis, for any spatial consine $u\in[-1,1]$ we define
the unit-norm 1D DFT vector
{\small\begin{equation}
\mathbf{a}_y(u)\triangleq
\frac{1}{\sqrt{N_y}}
\big[
e^{-j\frac{2\pi}{\lambda}du\delta_0},
e^{-j\frac{2\pi}{\lambda}du\delta_1},
\dots,
e^{-j\frac{2\pi}{\lambda}du\delta_{N_y-1}}
\big]^{\mathsf T},
\label{eq:1d_dft_ay}
\end{equation}}
and similarly along the $z$-axis
{\small\begin{equation}
\mathbf{a}_z(v)\triangleq
\frac{1}{\sqrt{N_z}}
\big[
e^{-j\frac{2\pi}{\lambda}dv\delta_0},
e^{-j\frac{2\pi}{\lambda}dv\delta_1},
\dots,
e^{-j\frac{2\pi}{\lambda}dv\delta_{N_z-1}}
\big]^{\mathsf T},
\label{eq:1d_dft_az}
\end{equation}}
Thus, a 2D DFT codeword on the UPA
pointing to $(u_n,v_m)$ is defined by the Kronecker product
$
\mathbf{a}(u_n,v_m)
\triangleq
\mathbf{a}_y(u_n)\otimes \mathbf{a}_z(v_m)\in\mathbb{C}^{N_yN_z}.
$

\subsection{Separable Fresnel approximation}
\noindent\hspace*{1em}Let $r^{(i,j)}$ denote the distance from element $(i,j)$ to a path at
$(\theta,\phi,r)$. A second-order (Fresnel) expansion around the array
center yields
{\small
\begin{align}
r^{(i,j)}-r
\approx
&-d\big(u\,\delta_i+v\,\delta_j\big)
+\frac{d^2}{2r}\Big((1-u^2)\delta_i^2+(1-v^2)\delta_j^2\Big)\nonumber\\
&-\frac{d^2}{r}\,u v\,\delta_i\delta_j.
\label{eq:fresnel_upa_full}
\end{align}}
Following standard large-array UPA codebook analyses \cite{wuMultipleAccessNearField2023},
the mixed quadratic term $-\tfrac{d^2}{r}uv\,\delta_i\delta_j$ is much
smaller than the dominant axis-wise terms over the operational near-field
region and can be safely neglected. The resulting near-field phase thus
approximately separates along the $y$- and $z$-axes, and the steering
vector becomes separable:
\begin{equation}
\mathbf{b}(u,v,r) ~\approx~ \mathbf{b}_y(u,r)\otimes \mathbf{b}_z(v,r),
\label{eq:upa_separable_sv}
\end{equation}
where
{\footnotesize
\begin{align}
\big[\mathbf{b}_y(u,r)\big]_i
&=\frac{1}{\sqrt{N_y}}
\exp\!\Big\{-j\frac{2\pi}{\lambda}\Big(-d\,u\,\delta_i+\frac{d^2}{2r}(1-u^2)\delta_i^2\Big)\Big\},
\\
\big[\mathbf{b}_z(v,r)\big]_j
&=\frac{1}{\sqrt{N_z}}
\exp\!\Big\{-j\frac{2\pi}{\lambda}\Big(-d\,v\,\delta_j+\frac{d^2}{2r}(1-v^2)\delta_j^2\Big)\Big\}.
\label{eq:upa_separable_sv_uv}
\end{align}}

\subsection{2D beam-pattern factorization and plateau support}
\noindent\hspace*{1em}Let one of the true paths be $(u_0,v_0,r_0)$. Under the separable Fresnel model
\Cref{eq:upa_separable_sv}, the beam pattern between the near-field
steering vector and a 2D DFT codeword factorizes as
\begin{align}
&G(u_0,v_0,r_0;u_n,v_m)\nonumber\\
&\triangleq \big|\mathbf{b}(u_0,v_0,r_0)^{H}\mathbf{a}(u_n,v_m)\big|\nonumber\\
&\approx
\big|\mathbf{b}_y(u_0,r_0)^{H}\mathbf{a}_y(u_n)\big|\;
\big|\mathbf{b}_z(v_0,r_0)^{H}\mathbf{a}_z(v_m)\big|.
\label{eq:2d_factorization}
\end{align}
Hence, the 2D magnitude response is the product of two 1D responses.

We now separately analyse the beam pattern for 2 directions. To quantify the 1D spreading, define the axis-wise normalized responses for z-axis
\begin{align}
G_{z,\mathrm{norm}}(v_0,r_0;v)
&\triangleq
\frac{\big|\mathbf{b}_z(v_0,r_0)^{H}\mathbf{a}_z(v)\big|}
{\big|\mathbf{b}_z(v_0,r_0)^{H}\mathbf{a}_z(v_0)\big|}.
\label{eq:axis_norm_resp}
\end{align}
The corresponding 6-dB lobe sets are
\begin{align}
\Phi^{z}_{6\mathrm{dB}}(v_0,r_0)
&\triangleq
\bigl\{v\mid\,G_{z,\mathrm{norm}}(v_0,r_0;v)\ge 1/2\bigr\},
\end{align}
and the associated widths are
\begin{align}
&B_z(v_0,r_0)\triangleq \max(\Phi^{z}_{6\mathrm{dB}})-\min(\Phi^{z}_{6\mathrm{dB}}).
\label{eq:axis_width_def}
\end{align}
For y-axis the definition is the same.

In the near-field ULA analysis \cite{wang2025lowcomplexitynearfieldbeamtraining},
the 6-dB lobe width for a path at
$(\xi_0,r_0)$ has the closed-form as
\begin{equation}
B(\xi_0,r_0)=Nd\,\frac{1-\xi_0^2}{r_0}.
\label{eq:beamwidth_ula}
\end{equation}
In our UPA setting, for a given path located at $(u_0,v_0,r_0)$, the separable Fresnel
model in \Cref{eq:upa_separable_sv_uv} shows that $\mathbf{b}_y(u_0,r_0)$ and
$\mathbf{b}_z(v_0,r_0)$ have the same quadratic-phase structure as the near-field ULA
steering vector in \cite{wang2025lowcomplexitynearfieldbeamtraining}, with apertures
$N_y d$ and $N_z d$. The same 6-dB width law applies along each axis, yielding the per-axis
spreading widths around the true spatial variables $u_0$ and $v_0$:
\begin{equation}
B_y(u_0,r_0)=N_y d\,\frac{1-u_0^2}{r_0},\;
B_z(v_0,r_0)=N_z d\,\frac{1-v_0^2}{r_0}.
\label{eq:beamwidth_upa_uv}
\end{equation}
These widths quantify how far in the continuous $(u,v)$ domain the near-field energy of
this path spreads from its peak along each axis due to the near-field quadratic phase.
Combined with the factorization in \Cref{eq:2d_factorization}, a convenient and support approximation is given by the Cartesian product of the two axis-wise
6-dB sets, which matches the
connected plateau with sharp edges observed in DFT beamspace
\cite{wang2025lowcomplexitynearfieldbeamtraining,Cui_and_Dai_channel_model}.

\begin{figure}[!t]
\centering
\includegraphics[width=0.8\linewidth]{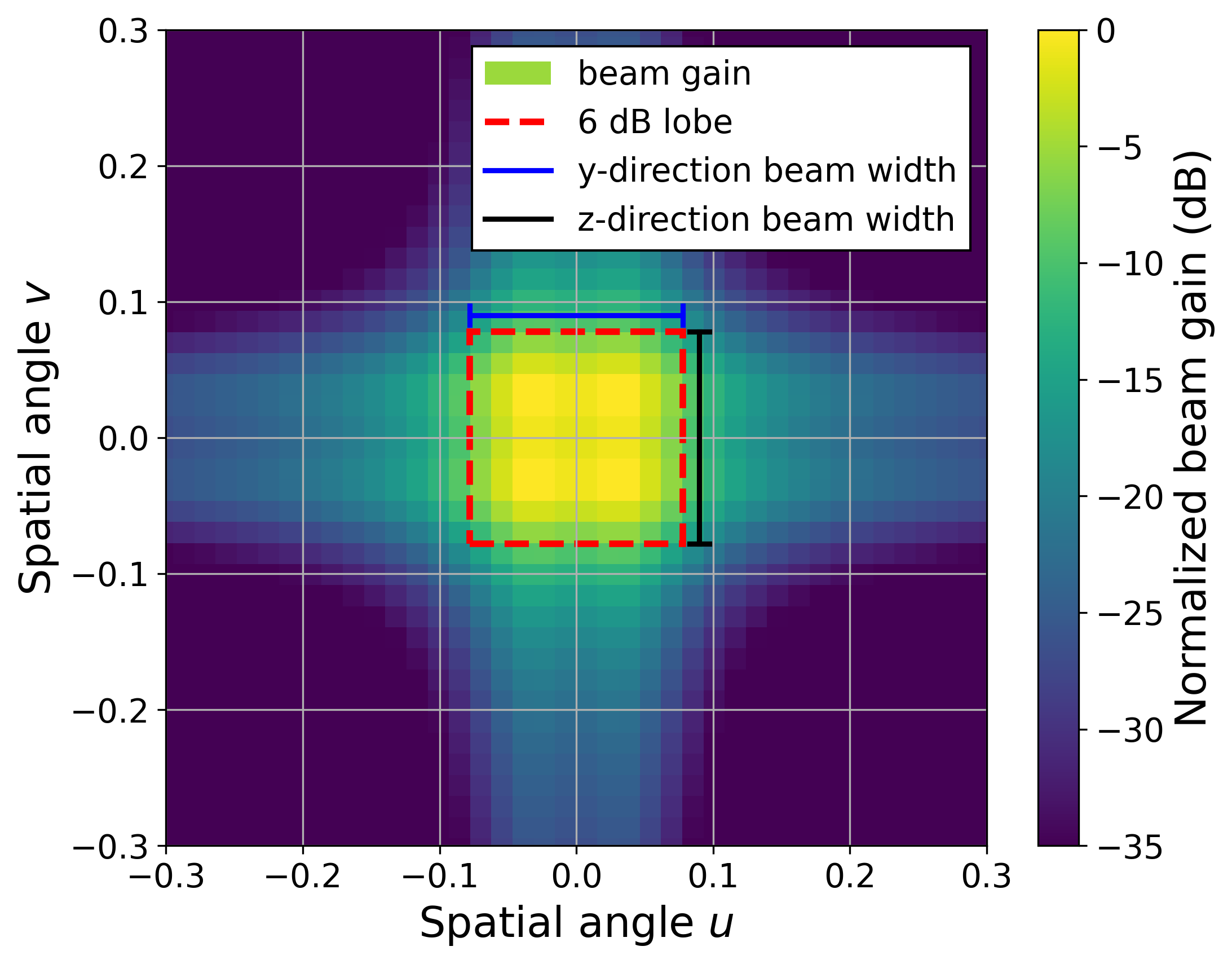}
\caption{Normalized DFT--beamspace magnitude on a $(128\times 128)$ UPA at
$f_c=28~\mathrm{GHz}$.}
\label{fig:upa_mainlobe}
\end{figure}

\begin{figure}[t]
    \centering
    % (a) NMSE vs SNR
    \subfloat[\tiny(c)][\textrm{\footnotesize u-cut (y-direction beam width)}]{
        \includegraphics[width=0.46\linewidth]{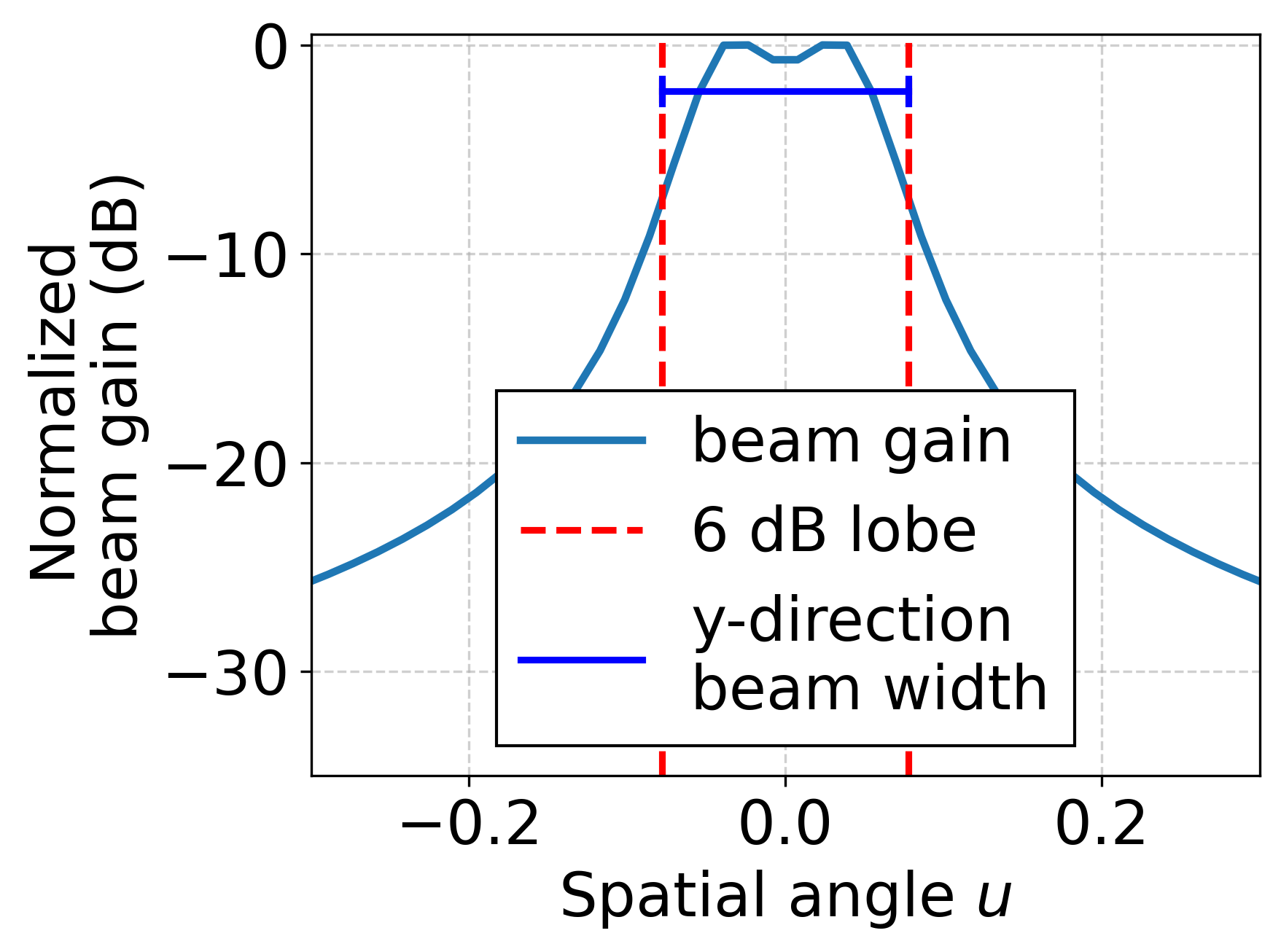}%
        \label{fig:upa_cut_u}
    }
    \hfill
    % (b) Correlation vs SNR
    \subfloat[\tiny(c)][\textrm{\footnotesize v-cut (z-direction beam width)}]{
        \includegraphics[width=0.46\linewidth]{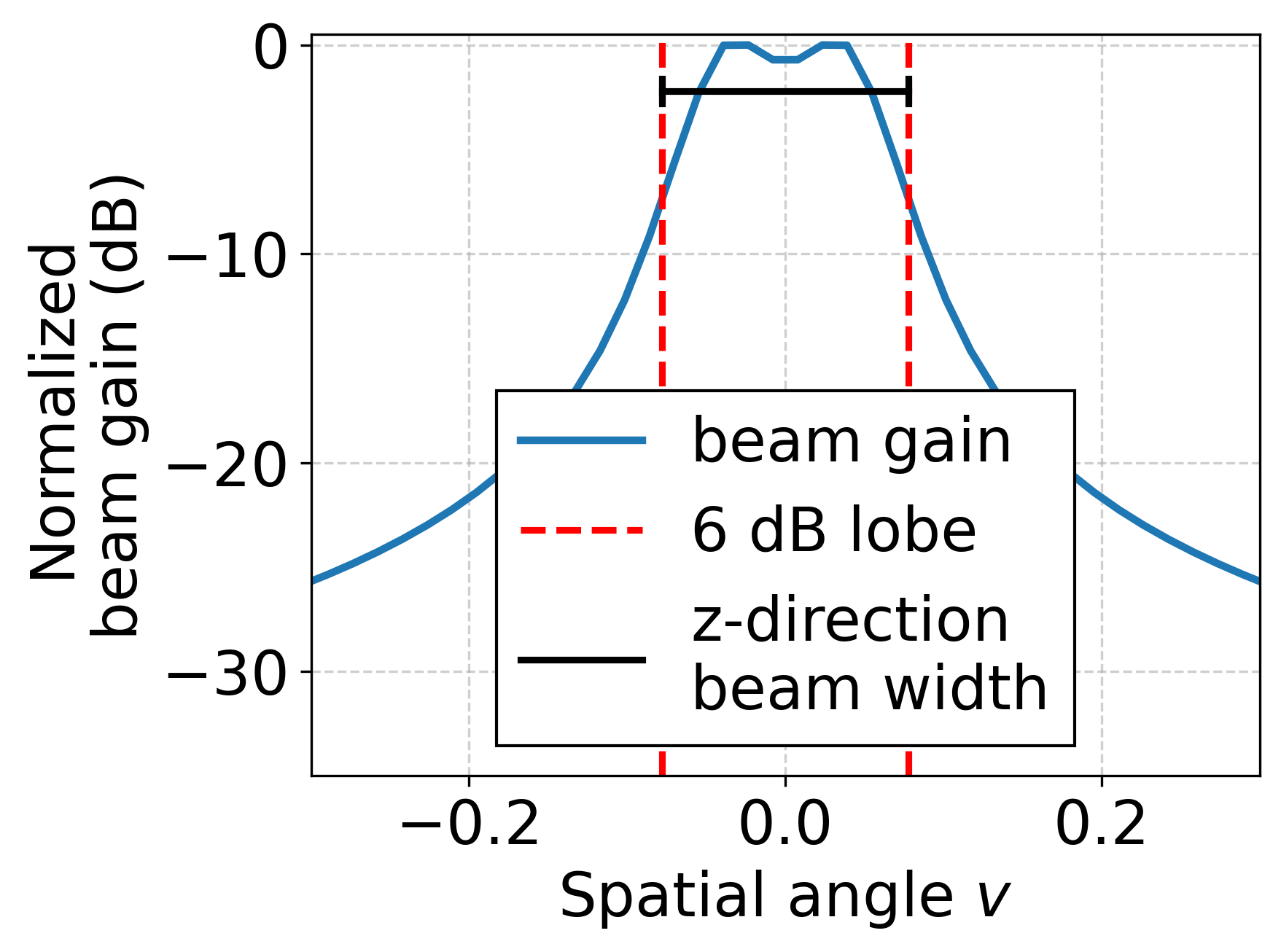}%
        \label{fig:upa_cut_v}
    }
    \caption{
        1D cuts of the normalized gain and the 6-dB lobe widths.
    }
    \label{fig:upa_main_lobe2}
\end{figure}
An example of the UPA beam pattern is shown in \Cref{fig:upa_mainlobe}.
The red dashed rectangle marks the 6-dB lobe region, defined as the
set of DFT beam indices whose magnitude exceeds $0.5$ of the central peak, and
highlights the compact plateau induced by near-field effect. \Cref{fig:upa_main_lobe2} further shows the 1D $u$- and $v$-cuts through the peak of \Cref{fig:upa_mainlobe}; the 6-dB thresholds (red dashed lines) delineate the per-axis lobe widths $B_y$ and $B_z$, confirming the flat-top plateau behavior.

\subsection{Effective sparsity level in 2D DFT beamspace}
\noindent\hspace*{1em}Let $\Delta_y$ and $\Delta_z$ denote the DFT grid spacings along the
$u$- and $v$-axes, respectively. Approximating each path's 2D
lobe support by a rectangle with side lengths $B_y$ and $B_z$, the
number of active DFT coefficients per path is approximately
$B_yB_z/(\Delta_y\Delta_z)$. For $L$ paths with non-overlapping lobe
supports (the worst case), the expected sparsity level $K_{\mathrm{UPA}}$ (i.e., the expected number of active
DFT coefficients) scales as
\begin{equation}
\mathbb{E}[K_{\mathrm{UPA}}]
~\approx~
\frac{LN_yN_zd^2}{\Delta_y\Delta_z}\,
\mathbb{E}\!\left[\frac{(1-u^2)(1-v^2)}{r^2}\right].
\label{eq:sparsity_upa_general}
\end{equation}
Note that $u$ and $v$ are coupled through $u=\sin\theta\sin\phi=\sqrt{1-v^2}\,s$ with
$s\triangleq \sin\phi$. For typical near-field user distributions, the expectation in
\Cref{eq:sparsity_upa_general} can be evaluated in closed form. For a $N_y\times N_z=128\times 16$ UPA at $f_c=28$~GHz with $d=\lambda/2$
and the standard 2D DFT grid $\Delta_y=2/N_y$ and $\Delta_z=2/N_z$, consider
$
v\sim \mathcal{U}\!\left[-\tfrac{1}{2},\tfrac{1}{2}\right],\;
s\sim \mathcal{U}\!\left[-\tfrac{1}{2},\tfrac{1}{2}\right],\;
r\sim \mathcal{U}\!\left[r_{\mathrm{F}},\frac{r_{\mathrm{R}}}{20}\right].
$
In this case, for an $L=5$-path channel, \Cref{eq:sparsity_upa_general} yields
$
\mathbb{E}[K_{\mathrm{UPA}}]\approx 8.6,
$
confirming that the near-field channel remains strongly sparse in the 2D DFT beamspace under this setting.

This analysis suggests that the near-field 2D DFT coefficients are not only sparse but also
clustered into compact, plateau-like lobe supports with sharp boundaries.
A pure LASSO tends to induce amplitude bias and favors scattered nonzeros,
which can erode flat lobe levels and break a contiguous block into fragmented patches.
In contrast, TV regularization promotes piecewise-constant structure
within each lobe while preserving edge discontinuities, making it a natural complement to
LASSO for denoising near-field DFT beamspace patterns.

\section{Compressive beam training}\label{sec:ulamethod}
\noindent\hspace*{1em}We present a multi-stage compressive beam training method motivated by the
near-field DFT beam patterns in \Cref{sec:beampattern}. Consider an
$N_y\times N_z$ UPA and denote the column-stacked array-domain channel by
$\mathbf{h}\in\mathbb{C}^{N}$ with $N\triangleq N_yN_z$.
Define the unitary 2D DFT basis without oversampling as
$
\mathbf{F}\in\mathbb{C}^{N\times N}.
$
We represent the channel in the 2D DFT beamspace as
$
\mathbf{h}=\mathbf{F}\mathbf{s},
$
where $\mathbf{s}\in\mathbb{C}^{N}$ is the beamspace coefficient vector.
To expose the 2D structure, we reshape $\mathbf{s}$ into a matrix
$S\in\mathbb{C}^{N_y\times N_z}$ using the same column-stacking order, i.e.,
$\mathbf{s}=\mathrm{vec}(S)$.

During training, the BS transmits $M$ pilot beams $\{\mathbf{v}_m\}_{m=1}^{M}$
with $M\ll N$, each generated as a masked 2D-DFT combination
$
\mathbf{v}_m=\mathbf{F}\,\mathbf{p}_m,\;
\mathbf{p}_m\sim\mathcal{CN}(\mathbf{0},\tfrac{1}{N}\mathbf{I}_N).
$
With pilot $x=1$, the received observation is
$y_m=\mathbf{h}^{H}\mathbf{v}_m+w_m=\mathbf{s}^{H}\mathbf{p}_m+w_m$,
where $w_m\sim\mathcal{CN}(0,\sigma^2)$.
Stacking $\{y_m\}$ yields the linear beamspace model
\begin{equation}
\mathbf{y}=\boldsymbol{\Phi}\,\mathbf{s}+\mathbf{w},\;
\boldsymbol{\Phi}\triangleq
\begin{bmatrix}
\mathbf{p}_1^{H}\\ \vdots \\ \mathbf{p}_M^{H}
\end{bmatrix}\in\mathbb{C}^{M\times N}.
\label{eq:sensing_model_upa}
\end{equation}

As shown in \Cref{sec:beampattern}, a near-field path produces a compact,
plateau-like lobe in the 2D DFT beamspace. This motivates a three-stage beam training scheme.

\subsubsection{Stage~I: support detection via LASSO}
We first obtain a coarse estimate by solving
\begin{equation}
\widehat{\mathbf{s}}^{(\ell_1)} \in
\arg\min_{\mathbf{s}\in\mathbb{C}^{N}}\,
\frac{1}{2}\|\mathbf{y}-\boldsymbol{\Phi}\mathbf{s}\|_{2}^{2}
+\lambda_{1}\|\mathbf{s}\|_{1}.
\label{eq:lasso_upa}
\end{equation}
Reshape $\widehat{\mathbf{s}}^{(\ell_1)}$ into
$\widehat{S}^{(\ell_1)}\in\mathbb{C}^{N_y\times N_z}$ via the
$\mathrm{vec}(\cdot)$ order, and form an initial support mask by top--$k$
selection on magnitudes:
\begin{equation}
\mathcal{M}_{0}=\operatorname*{Top}\!-\!k\big(|\widehat{S}^{(\ell_1)}|\big)
\subseteq\{1,\ldots,N_y\}\times\{1,\ldots,N_z\}.
\label{eq:mask0_upa}
\end{equation}

\subsubsection{Stage~II: 2D support dilation}
To capture the full connected lobe support, we dilate $\mathcal{M}_0$ by a
2D radius $(r_y,r_z)\in\mathbb{N}^2$:
{\small\begin{equation}
\widetilde{\mathcal{M}}
=\Big\{(i,j)\mid \exists(i',j')\in\mathcal{M}_0\ \text{s.t.}\ |i-i'|\le r_y,\ |j-j'|\le r_z\Big\}.
\label{eq:dilation_upa}
\end{equation}}

\subsubsection{Stage~III: magnitude--TV refinement on the dilated mask}
We refine the beamspace coefficients on $\widetilde{\mathcal{M}}$ by
penalizing the 2D total variation of the magnitude map $|S|$:
\begin{align}
\widehat{S}\in \arg\min_{S\in\mathbb{C}^{N_y\times N_z}}
&\ \frac{1}{2}\big\|\mathbf{y}-\boldsymbol{\Phi}\,\mathrm{vec}(S)\big\|_{2}^{2}
+\lambda_{\mathrm{TV}}\, \mathrm{TV}_{2\mathrm{D}}\!\left(|S|;\,\widetilde{\mathcal{M}}\right)
\nonumber\\
\text{s.t.}\quad
&\ S_{i,j}=0,\ \ \forall (i,j)\notin \widetilde{\mathcal{M}}.
\label{eq:tv2d_problem_upa}
\end{align}
We use the anisotropic masked TV
{\small\begin{equation}
\mathrm{TV}_{2\mathrm{D}}\!\left(|S|;\,\widetilde{\mathcal{M}}\right)
=\!\!\sum_{\substack{(i,j)\in\widetilde{\mathcal{M}}}}\!\!\Bigl||S_{i+1,j}|-|S_{i,j}|\Bigr|
+\!\!\sum_{\substack{(i,j)\in\widetilde{\mathcal{M}}}}\!\!\Bigl||S_{i,j+1}|-|S_{i,j}|\Bigr|,
\label{eq:tv_2d_mask_upa}
\end{equation}}
where out-of-range terms are ignored at the boundaries. The final
beamspace estimate is $\widehat{\mathbf{s}}=\mathrm{vec}(\widehat{S})$. We recover the array-domain channel via
 $
\widehat{\mathbf{h}}=\mathbf{F}\,\widehat{\mathbf{s}}.
$
Finally, we form the beamforming vector by normalizing the recovered channel $
\widehat{\mathbf{v}}=\widehat{\mathbf{h}}/\|\widehat{\mathbf{h}}\|_{2}.
$

\section{Simulation Results}\label{sec:upa_results}
\begin{figure}[!t]
\captionsetup{justification=centering,singlelinecheck=true}
\centering
    \includegraphics[width=0.8\linewidth]{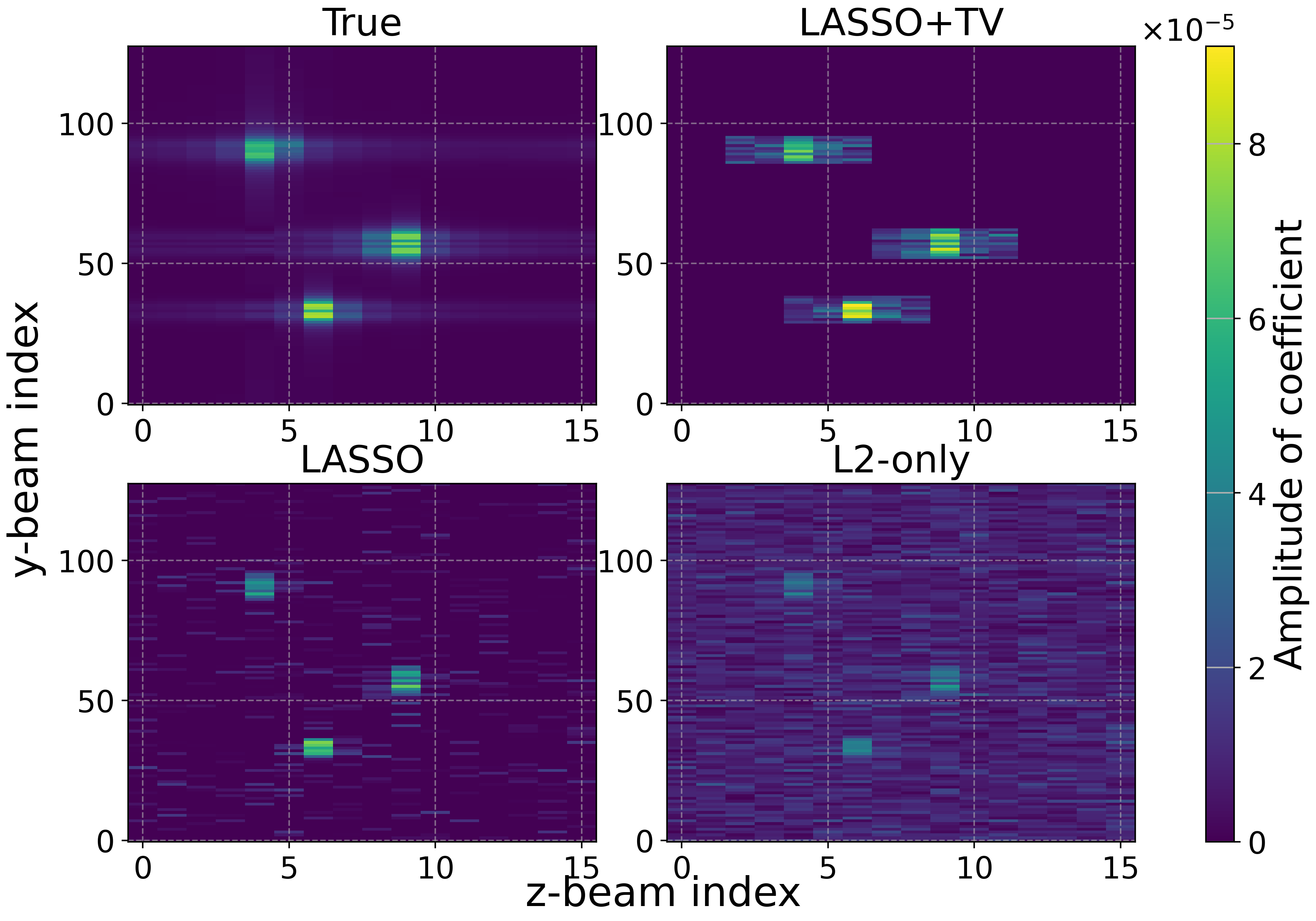}
% \captionsetup{justification=raggedright,singlelinecheck=false}
\caption{DFT beamspace magnitude on a UPA near-field channel.}
\label{fig:upa_beamspace}
\end{figure}

\noindent\hspace*{1em}We consider a single BS equipped with a UPA of \(N_y = 128,N_z=16\) antenna elements and the central frequency is $f_c=28$GHz. We first plot comparison of recovered channel coefficients power of different methods. We consider a multi-path channel with $L=3$ and SNR=$0$ dB. As shown in \Cref{fig:upa_beamspace}, the true beamspace exhibits a compact 2D plateau centered around each path, consistent with the separable near-field lobe structure discussed in \Cref{sec:beampattern}. The proposed LASSO+TV estimator closely matches both the extent and the height of this plateau while strongly suppressing out-of-support leakage. In contrast, pure LASSO produces a fragmented, ridge-like pattern with over-shrunk amplitudes and spurious side patches, whereas the L2-only baseline is dominated by diffuse noise spread over the entire beamspace. This proves that LASSO+TV restores a clean rectangular lobe block that aligns with the expected UPA near-field beam pattern, while LASSO and L2 fail to recover a clear 2D support.

\begin{figure}[t]
    \centering
    % (a) NMSE vs SNR
    \subfloat[\tiny(c)][\textrm{\footnotesize NMSE vs SNR}]{
        \includegraphics[width=0.46\linewidth]{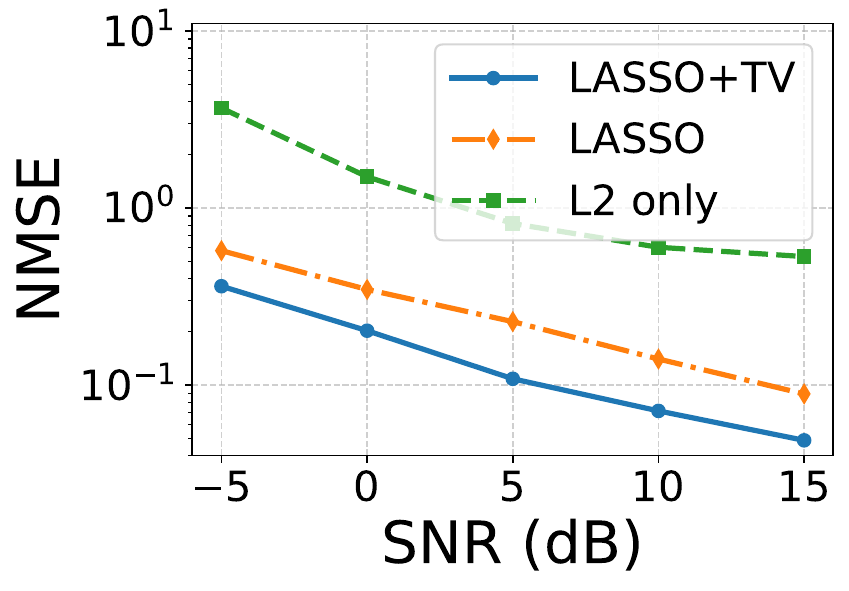}%
        \label{fig:upa_nmse_snr}
    }
    \hfill
    % (b) Correlation vs SNR
    \subfloat[\tiny(c)][\textrm{\footnotesize Normalized beamforming gain vs SNR}]{
        \includegraphics[width=0.46\linewidth]{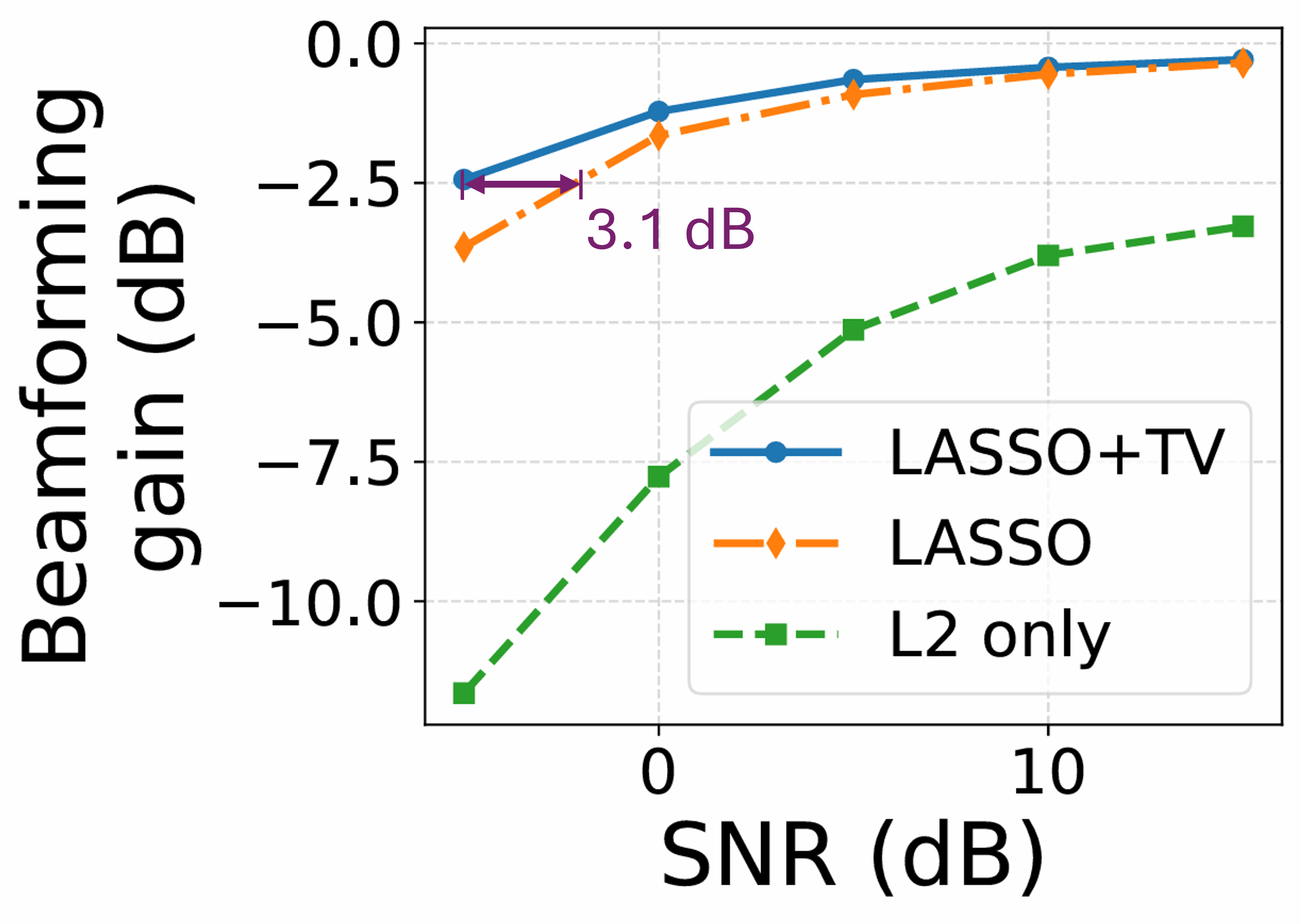}%
        \label{fig:upa_corr_snr}
    }
    \caption{
        Simulation results for the SNR sweeping.
    }
    \label{fig:upa_snr_results}
\end{figure}

We next consider Monte Carlo simulations with $N_{\mathrm{user}}=500$ independent user channels,
each modeled as an $L=5$-path near-field link with uniform random location
$(u,s,r)$ in the UPA near-field region. The NMSE and averaging
normalized beamforming gain are shown in \Cref{fig:upa_nmse_snr,fig:upa_corr_snr}.
 For each realization, we sweep
the downlink $\mathrm{SNR}$ from $-5$~dB to $15$~dB. The
per-realization NMSE and normalized beamforming gain(dB) are defined as
$
\mathrm{NMSE}(\widehat{\mathbf{h}},\mathbf{h})
\triangleq
\frac{\|\widehat{\mathbf{h}}-\mathbf{h}\|_{2}^{2}}{\|\mathbf{h}\|_{2}^{2}},
\;
\rho(\widehat{\mathbf{h}},\mathbf{h})
\triangleq
20\log_{10}\Big(\frac{|\widehat{\mathbf{h}}^{H}\mathbf{h}|}{\|\widehat{\mathbf{h}}\|_{2}\,\|\mathbf{h}\|_{2}}\Big).
$

From \Cref{fig:upa_nmse_snr}, the proposed LASSO+TV estimator consistently
achieves the lowest NMSE across the whole SNR range. At low SNR, it reduces
NMSE by about $16\%$ over LASSO and about $87\%$ over the L2-only baseline.
At medium SNR, the corresponding NMSE reductions are about $32\%$ and $84\%$,
respectively, and at high SNR the gains remain substantial (about $27\%$
over LASSO and $88\%$ over L2-only). 

The mean normalized beamforming gain in \Cref{fig:upa_corr_snr} 
further demonstrates a consistent advantage of LASSO+TV. In particular, at
$\mathrm{SNR}=0$~dB, LASSO+TV achieves a gain of $-1.22$~dB, which is about
$0.43$~dB higher than plain LASSO and about $6.55$~dB higher than L2-only. At
$\mathrm{SNR}=15$~dB, LASSO+TV improves to $-0.29$~dB, remaining about $0.06$~dB above
LASSO and about $2.99$~dB above L2-only. Overall, at the $-2.5$~dB beamforming-gain level,
LASSO+TV provides an effective SNR gain of about $3.1$~dB over plain LASSO, confirming
that enforcing both sparsity and plateau-preserving local smoothness in the DFT beamspace
translates into robust beamforming gains in the UPA setting.

\begin{figure}[t]
\centering
\subfloat[\tiny(a)][\textrm{\footnotesize NMSE vs measurement factor}]{
    \includegraphics[width=0.46\linewidth]{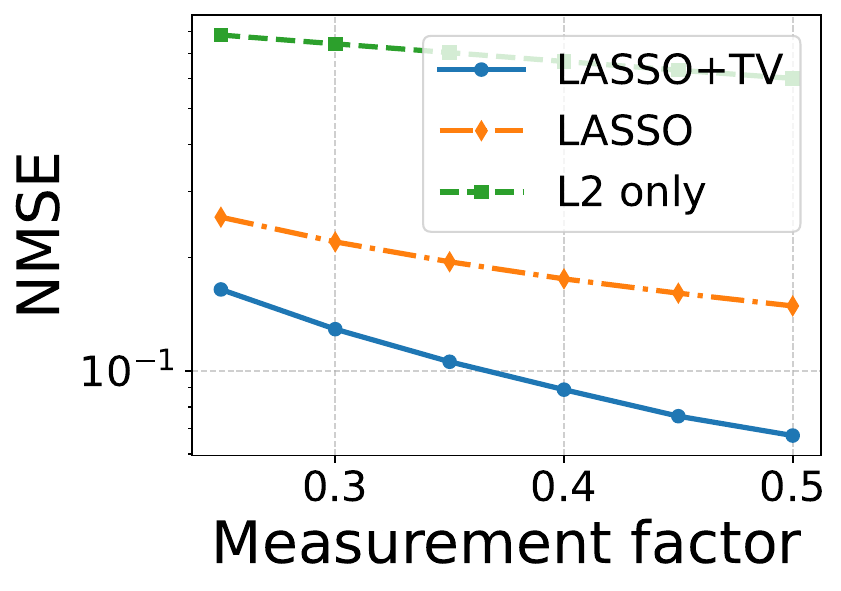}%
    \label{fig:upa_nmse_meas}
}
\hfill
\subfloat[\tiny(b)][\textrm{\footnotesize NMSE vs number of path}]{
    \includegraphics[width=0.46\linewidth]{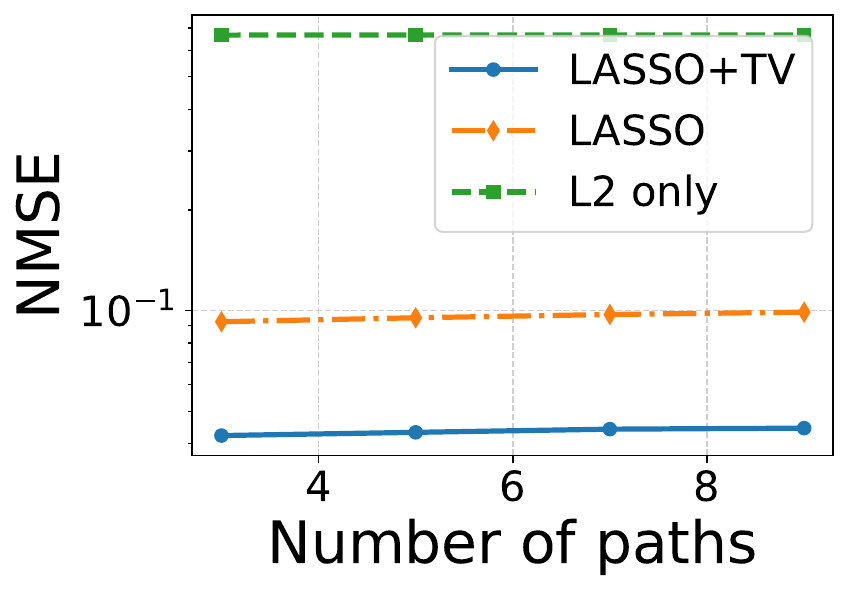}%
    \label{fig:upa_path_num}
}
\caption{Simulation results for the pilot-budget and path-number sweeping.}
\label{fig:upa_meas_sweep}
\end{figure}

We next study the impact of pilot budget under the same UPA setting and set $\mathrm{SNR}=10$~dB.
The measurement factor $M/N$ is swept from $0.25$ to $0.50$, corresponding to $M=512$--$1024$ pilots. From \Cref{fig:upa_nmse_meas}, increasing $M/N$ improves all methods, and LASSO+TV is uniformly best.
At $M/N=0.25$, LASSO+TV attains $\mathrm{NMSE}=2.19\times10^{-1}$, which is lower than LASSO by $3.69\times10^{-2}$, corresponding to a $14\%$ NMSE reduction, and lower than L2-only by $5.64\times10^{-1}$, corresponding to a $72\%$ NMSE reduction.
At $M/N=0.50$, LASSO+TV reduces NMSE by $40\%$ relative to LASSO, and by  $85\%$ relative to L2-only.

In \Cref{fig:upa_path_num}, we evaluate the impact of the number of paths $L$ on channel recovery performance. As $L$ increases, all methods exhibit a slight degradation, consistent with the reduced sparsity of the DFT-beamspace representation under richer multipath. Nevertheless, the proposed LASSO+TV remains consistently superior to plain LASSO across the entire range of $L$, providing a stable NMSE reduction of roughly 35\%-40\%. This highlights the necessity of exploiting beamspace sparsity and plateau structure for reliable near-field recovery.

% The beamforming-gain trends in \Cref{fig:upa_bfgain_meas} are consistent.
% At $M/N=0.25$, LASSO+TV provides a mean gain of $-1.04$~dB, exceeding LASSO by $0.06$~dB and L2-only by $5.53$~dB.
% At $M/N=0.50$, the gain improves to $-0.40$~dB, maintaining a $0.18$~dB advantage over LASSO and a $3.41$~dB advantage over L2-only.
% Overall, the proposed plateau-preserving refinement translates into consistent NMSE and beamforming-gain improvements across the pilot-budget range.

% \begin{figure}[t]
% \centering
%     \includegraphics[width=0.5\linewidth]{pics/mc_results_meas_sweep_SNR15_L5_500users_upa2_mc_nmse_vs_measfactor.pdf}%
%     % \label{fig:upa_nmse_meas}
% \caption{Simulation results for the pilot-budget sweeping.}
% \label{fig:upa_meas_sweep}
% \end{figure}

\section{Conclusion}\label{sec:clu}
\noindent\hspace*{1em}In this paper, we studied near-field beam training for multiple-path scenarios
using conventional DFT codebooks and low-overhead Gaussian-masked DFT pilots.
Motivated by the plateau-like near-field DFT beam patterns, we proposed a
LASSO--dilation--magnitude-TV refinement that exploits clustered sparsity and
preserves lobe edges without designing polar-domain codebooks. Simulations for UPA near-field multipath channels show consistent NMSE and
beamforming-correlation gains over least squares and plain LASSO.
In the considered UPA setting, the proposed method achieves about a $2$--$3$~dB effective
SNR gain over LASSO.

\printbibliography

\vfill

\end{document}